# Scientific and Technological News Recommendation Based on Knowledge Graph with User Perception

Yuyao Zeng, Junping Du*, Zhe Xue , Ang Li

School of Computer Science, Beijing Key Laboratory of Intelligent Telecommunication Software and Multimedia, Beijing University of Posts and Telecommunications, Beijing 100876, China

**Abstract:** Existing research usually utilizes side information such as social network or item attributes to improve the performance of collaborative filtering-based recommender systems. In this paper, the knowledge graph with user perception is used to acquire the source of side information. We proposed KGUPN to address the limitations of existing embedding-based and path-based knowledge graph-aware recommendation methods, an end-to-end framework that integrates knowledge graph and user awareness into scientific and technological news recommendation systems. KGUPN contains three main layers, which are the propagation representation layer, the contextual information layer and collaborative relation layer. The propagation representation layer improves the representation of an entity by recursively propagating embeddings from its neighbors (which can be users, news, or relationships) in the knowledge graph. The contextual information layer improves the representation of entities by encoding the behavioral information of entities appearing in the news. The collaborative relation layer complements the relationship between entities in the news knowledge graph. Experimental results on real-world datasets show that KGUPN significantly outperforms state-of-the-art baselines in scientific and technological news recommendation.

**Keywords:** Recommendation; Knowledge graph; User perception

## 1 Introduction

With the development of the World Wide Web, online news platforms such as Google News, microblogs [1] and Microsoft News emerge one after another. Due to the convenience and speed of online news, people's news reading habits have gradually shifted from traditional media such as newspapers and TV to the Internet. Tech News follows the latest developments in technology. The latest scientific and technological information [2] is reported in real time, which makes technology news a popular and indispensable type of news. News websites collect news from various sources, which makes the number of news articles grow exponentially. At the same time, because of its rich semantics, short timeliness, and many types of technology news, it leads to problems such as user information overload. In order to help users quickly browse the news they are interested in and improve the reading experience; personalized news recommendation technology came into being.

Traditional news recommendation methods include methods based on collaborative filtering [3-7] , content-based methods [8][9], and hybrid methods [10][11], which generate user and item features from interaction matrices. For example, in scoring-related recommender systems, the interaction between users and items usually adopts collaborative filtering [12][13]. However, the special challenges faced by technology news recommendation make traditional recommendation algorithms less effective. First, technology news is updated very fast on online news platforms and is highly time-sensitive, with the release of constantly updated technology news, the existing technology news will lose its timeliness, and the correlation between news will also be invalid. Therefore, cold start is also an important problem to be solved in the technology news recommendation system. Second, users usually do not rate news, and how to mine user interests from user clicks and historical recommendations is an urgent problem to be solved in news recommendation. Third, the titles and texts of science and technology news contain a large amount of rich text information, which can be parsed into many knowledge entities and common sense, and news recommendation is made through the correlation of knowledge entities and common-sense reasoning, but the existing recommendation algorithms only pass ID to simply represent text information, it is difficult to find the correlation between texts. This also causes problems such as homogeneity of recommended content.

To solve the existing challenges of technology news recommendation mentioned above, in this paper, we propose a new framework for technology news recommendation using knowledge graphs and user portraits, namely the knowledge graph user perception network (KGUPN). KGUPN automatically mines the higher-order connection relationships in KG along the links in the knowledge graph, iteratively expands the potential interests of users, and introduces the context information and entity collaborative relationships of science and technology news entities, thereby establishing a hybrid structure of KG and user-item graphs, stimulating user preferences propagated to the knowledge entity set. With the support of the user's historical click items, the user's preference distribution for candidate items is formed, which can be used to predict the final click probability of science and technology news.

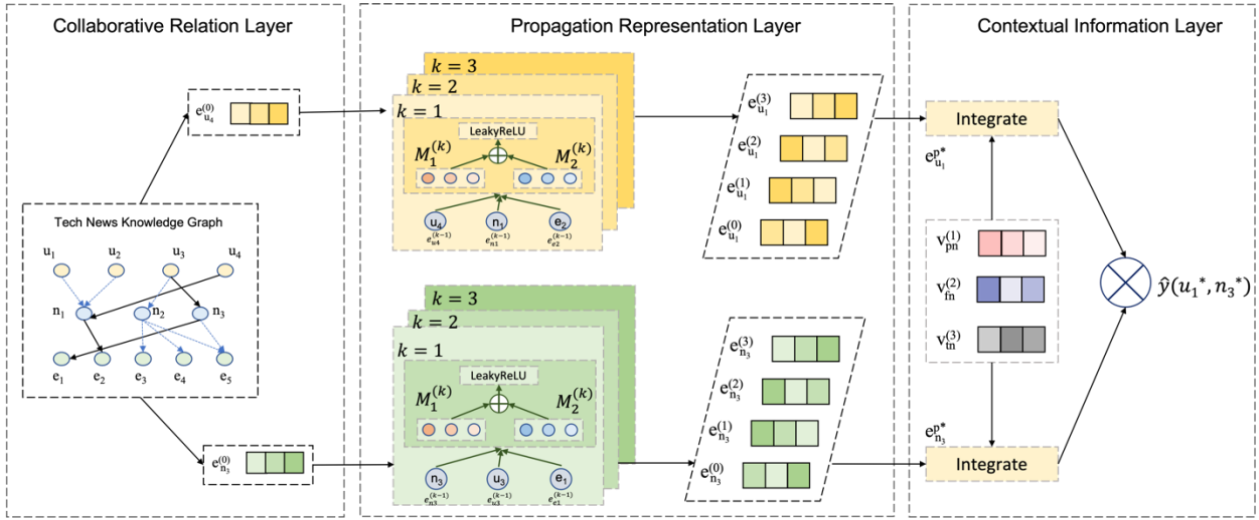

**Figure 1** An overview of the proposed KGUPN mode

In summary, the main contributions of our work are as follows:

- We propose KGUPN, an end-to-end framework that utilizes knowledge graphs with user perception to assist scientific and technological news recommend systems. KGUPN utilizes collaborative relations and discovers users' hierarchical latent interests by iteratively propagating users' preferences in the KG.

- To fully utilize knowledge information, we propose three key layers in KGUPN, including a collaborative relations layer, a propagation representation layer, and a contextual information layer. Through ablation studies we verify that indeed each component contributes to the model.

- We conduct experiments on two real-world news recommendation scenarios and a benchmark dataset widely used for general recommendation, and the results demonstrate the effectiveness of KGUPN on several state-of-the-art baselines.

## 2 Related Work

### 2.1 News Recommendation System

Traditional news recommendation methods include methods based on collaborative filtering [3][4][5], content-based methods [8][9] and hybrid methods [10][11]. But collaborative filtering-based methods often suffer from cold-start problems because news items are often replaced. Content-based methods can alleviate the cold-start problem by analyzing the content of the news users browse to recommend similar news to users. However, these methods ignore the sequential information in the user's browsing history, making it difficult to learn users' changing interests.

Previous news recommendation works extract features from news items manually [12] or extract latent representations through neural models [14]. These methods ignore the importance of entities in the article. In the direction of integrating knowledge graphs for news recommendation, the most relevant work is DKN [15][16]. However, DKN only takes news headlines as input. While it is possible to expand to incorporate news organizations, this would lead to inefficiencies.

### 2.2 Graph Based Recommendation System

Existing knowledge graph-based recommendation algorithms can be roughly divided into two categories: Path-based schemes and Embedding-based schemes. Path-based approach combined with knowledge graph in the field of recommendation is mainly to select and construct paths of different patterns between entities by defining meta-paths on the knowledge graph [17][18] or a path selection algorithm [19][20], to mine various associations between users and items on the knowledge graph, and then realize recommendation prediction. Embedding-based schemes [21][22] and tracking algorithms [23] are mostly based on knowledge graph embedding algorithm. With the development of graph convolutional network [24][25], researchers try to use it to the topological structure information realizes modeling [26][27][28], takes the knowledge graph topology and recommendation prediction as multiple learning objectives, and uses the attention mechanism [29][30] to learn the neighborhood weights to obtain the embedded representation of users and items.

Existing works usually directly use general knowledge graphs [31-34]. In this work, we construct a science and technology news knowledge graph based on the collaborative relationship between science and technology news entities and the interaction between science and technology news and users. The knowledge graph we build is more specialized and incorporates user news interaction information.

## 3 Knowledge Graph User Perception Network

We propose a Knowledge Graph User Perception Network model for news (KGUPN), which can be used for science and technology news recommendation. Figure 1 shows the overall KGUPN framework, which consists of three key layers: a collaborative relations layer, a propagation representation layer and a contextual information layer.

### 3.1 Collaborative Relations Layer.

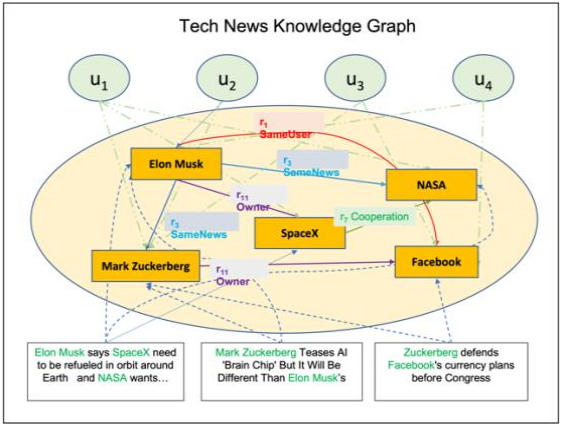

**Figure 2** The user perception knowledge graph with collaborative relations

The rich knowledge in knowledge graph can solve the problems of data discreteness and interpretability, therefore, in this paper, we mine the correlation of entities contained in news content and user clicks as supplementary knowledge of the KG. Based on the KG we built with Microsoft Satori; we supplement the correlation between entities in the knowledge graph. The correlation of newly added entities in KG includes two types, in the same news and click by the same user. The updated KG example diagram is shown in Figure 2.

**In the same news.** When two entities frequently appear in the same news, it often means that there is deep mutual relationship between the two entities. For example, Elon Musk and NASA often appear in the same news because they have the same scientific research goals. Such frequently co-occurring relations in the same news can be used for the mining and representation of deep relations in KG. Therefore, we add this relation to the KG as a complementary relation, such as "r3 SameNews" in Figure1.

**Clicked by the same user News.** entities that have been clicked by the same user can represent the interest correlation between entities. If multiple users have clicked on two entities at the same time, there may be some potential connection between the two entities, so a user who has clicked a certain entity may also be interested in the other news entity, even if the two entities do not have any direct relationship in the general knowledge graph. Therefore, we also add this relation to the KG as a supplementary relation, such as "r1 SameUser" in Fig.1

### 3.2 Propagation Representation Layer

We assume that entities in science and technology news and user-news interactions can be linked to a knowledge graph. A knowledge graph consists of a series of entity-relationship-entity triples, which can be expressed as $G = \{(u, r, n) \mid u \in U, r \in R, n \in N\}$ where $U$ denotes the set of user entities, $R$ represents the set of relations, and $N$ represents the set of science and technology news entities. $(u, r, n)$ represents that there is a relation $r$ from $u$ to $n$.

In addition, the entities and related users in the news article are represented as embedding vectors. A news entity $n$ is represented as an embedding vector $e_n \in R^d$, and a user $u$ is represented as $e_u \in R^d$, where $d$ represents the embedding size. We use detailed representations of news and mining of user-news higher-order relations to improve these embeddings. This approach leads to more efficient embeddings for news recommendation.

Considering that an entity is not only represented by its own embeddings, but can also be partially represented by its neighbors, we leverage the propagation structure of nodes and relations in the knowledge graph to refine the embeddings of users and news.

Directly interacting technology news (users) can most directly reflect the characteristics of users (technology news), users' historical clicks can reflect user preferences, users who have browsed technology news will be associated with this technology news and can also be used as a feature for technology news. We associate related users and news and exploit propagation to mine their potential relationships.

We use $e_u^{(k)}$ to represent the K-hop propagation of embedding of user $u$. This high-order connectivity is very helpful for inferring the deep connection between users and news, and this latent relationship can also be used to estimate user-news correlations.

By stacking k embedding propagation layers, users (and news) can receive messages propagated from their k-hop neighbors.

$$e_u^{(k)} = \text{LeakyReLU}\left(I_{u \leftarrow u}^{(k)} + \sum_{n \in A_u} I_{u \leftarrow n}^{(k)}\right) \quad (1)$$

Where $I_{u \leftarrow u}^{(k)}$ and $I_{u \leftarrow n}^{(k)}$ are defined as:

$$I_{u \leftarrow n}^{(k)} = d_{un}\left(M_1^{(k)} e_n^{(k-1)} + M_2^{(k)}\left(e_n^{(k-1)} \circ e_u^{(k-1)}\right)\right) \quad (2)$$

$$I_{u \leftarrow u}^{(k)} = M_1^{(k)} e_u^{(k-1)} \quad (3)$$

Where $M_1, M_2 \in R^{d_k \times d_{k-1}}$ are trainable transformation matrix, $d_k$ and $d_{k-1}$ are the transformation size. $e_n^{(k-1)}$ is the news representation generated from k-1 hop neighbors of user $u$, then it can be used to denotes the embedding of user $u$ at layer $k$ Deep relations are injected into the representation learning process by stacking multiple embedding propagation layers.

After K-hops of propagation, a set of representations for user u can be obtained, namely $\{e_u^{(1)} ... e_u^{(k-1)}, e_u^{(k)}\}$. In order to better utilize the user representations propagated from the K layers and concatenating the depth relationship information, we integrate them into the final embedding representation of the user $e_u^{p*}$. The final embedding of the news $e_n^{p*}$ is also obtained in the same way:

$$e_u^{p*} = e_u^{(0)} \diamond ... \diamond e_u^{(k-1)} \diamond e_u^{(k)} \quad (4)$$

$$e_n^{p*} = e_n^{(0)} \diamond ... \diamond e_n^{(k-1)} \diamond e_n^{(k)} \quad (5)$$

Where $\diamond$ is the concatenation operation. By doing so, we not only enrich the initial embeddings, but also allow the propagation range to be controlled by adjusting K.

### 3.3 Contextual Information Layer

The contextual relationship of an entity in the news affects the importance and relevance of the entity. To make the embedding of the entity describe the news more accurately, we design three contextual relation encodings to characterize the importance of entities: position, frequency, and category.

**Position Encoding.** Position encoding is used to represent where an entity appears, e.g., an entity that appears in both news headlines and body text is more important than an entity that appears only in the news body. We use a position vector $V_{pn}^{(1)}$ and combine it to the entity embedding, where $pn \in \{1,2\}$ denotes the news entity $e_n$ appears in title or body.

**Frequency Encoding.** Frequency represents the number of times an entity appears in the news and can be used as a measure of the importance of an entity. Therefore, we use matrix $V^{(2)}$ for frequency of each entity. We count the frequency of the appearance $fn$ for each entity, a frequency encoding vector is represented as $V_{fn}^{(2)}$, and combine it to the entity embedding. The upper limit of fn is set to 30.

**Category Encoding.** Entities in news can have a variety of categories, e.g., Elon Musk is a person, NASA is an organization, SpaceX is a company. We utilize a category matrix $V^{(3)}$. For each entity $i$ with category $tn$, then we combine this category encoding $V_{tn}^{(3)}$ to its embedding vector.

After the contextual embedding layer, for each entity n, its embedding vector as input for the next layer is a compound vector:

$$e_u^* = e_u^{p*} \oplus V_{pn}^{(1)} \oplus V_{fn}^{(2)} \oplus V_{tn}^{(3)} \quad (6)$$

Where $\oplus$ indicates the element-wise addition of vector.

Eventually, we conduct inner product of news and user embeddings, so the matching score is predicted as:

$$\hat{y}_{(u,n)} = (e_u^*)^T e_n^* \quad (7)$$

### 3.4 Loss Function

Assuming that observable interactions should be given better prediction values than unobserved ones, we use the pairwise BPR [35] loss to improve the recommendation model.

The learning algorithm of KGUPN is presented in Algorithm 1.

---
**Algorithm 1: Learning Algorithm for KGUPN**

**Input:** Tech News Knowledge Graph $\mathcal{G}$, number of epochs $E$, batch size $B$, number of layers $K$, learning rate $\lambda$

**Output:** Model's parameter set: $\Theta = \left\{M_1^{(k)}, M_2^{(k)}, e_u^{*(0)}, e_n^{*(0)}, \forall k \in \{1, \cdots, K\}\right\}$

1 Add user interaction and collaborative relations into $\mathcal{G}$;
2 Randomly initialize $e_u^{(0)}$ and $e_n^{(0)}$ from TransE;
3 **for** $e = 1$ to $E$ **do**
4    Generate the $e_{th}$ batch of size $B$ true and false triples from $\mathcal{G}$;
5    **for** $k = 1$ to $K - 1$ **do**
6       Calculate $e_u^{k+1}$ and $e_n^{k+1}$ by using Eq. (1)-(5);
7    **end**
8    Concatenate $[e_u^{(0)}, ... e_u^{(k-1)}, e_u^{(k)}]$ into $e_u^{p*}$, and concatenate $[e_n^{(0)}, ... e_n^{(k-1)}, e_n^{(k)}]$ into $e_n^{p*}$;
9    Calculate $e_u^*$ and $e_n^*$ by using Eq.(7);
10   Update $\Theta_{e+1}$ by gradient descent with learning rate $\lambda$;
11 **end**
12 **return** $\Theta_E$

---

## 4 Experiment

### 4.1 Research Questions

**RQ1:** Does our Knowledge Graph User Perception Network method KGUPN outperform the state-of-the-art baseline algorithms?

**RQ2:** How do different components settings (e.g., contextual embedding, information distillation, and user perception) affect KGUPN?

**RQ3:** Can KGUPN provide reasonable explanations about user preferences towards news?

### 4.2 Datasets

**Table I** Statistics of the datasets

| DataSet | | MIND | MovieLens |
|---|---|---|---|
| User-News Interaction | User | 46,342 | 5389 |
| | News | 61,013 | 2445 |
| | Interaction | 455,470 | 253,772 |
| Knowledge Graph | Entities | 399,687 | 100,384 |
| | Relations | 11 | 6 |
| | Triplets | 3,425,590 | 517,097 |

Table II Comparison of recommend performance on MIND and MovieLens datasets

| Model | MIND | | | MovieLens | | |
|---|---|---|---|---|---|---|
| | Recall@10 | Recall@20 | NDCG@10 | Recall@10 | Recall@20 | NDCG@10 |
| CFKG | 0.1024 | 0.1922 | 0.1697 | 0.0378 | 0.0723 | 0.0786 |
| CKE | 0.1092 | 0.1927 | 0.1755 | 0.0397 | 0.0736 | 0.0792 |
| FM | 0.1349 | 0.1978 | 0.1933 | 0.0412 | 0.0778 | 0.0799 |
| DKN | 0,.1353 | 0.1993 | 0.1952 | 0.0429 | 0.0791 | 0.0810 |
| RippleNet | 0.1435 | 0.2064 | 0.1990 | 0.0469 | 0.0829 | 0.0824 |
| GC-MC | 0.1514 | 0.2098 | 0.2124 | 0.0495 | 0.0870 | 0.0896 |
| LibFM | 0.1531 | 0.2113 | 0.2172 | 0.0453 | 0.0818 | 0.0859 |
| **KGUPN** | **0.1593** | **0.2166** | **0.2278** | **0.0522** | **0.0914** | **0.0943** |

We employ the processed Microsoft news recommendation dataset MIND, as well as a benchmark dataset frequently used in recommender systems: MovieLens, which is publicly accessible and differs in domain, size, and sparsity, to thoroughly assess the efficacy of the suggested algorithm above.

**MIND[36]**: This data set was gathered from the Microsoft news website's anonymized usage records. It includes click statistics and behavioral diaries from users who clicked on at least five news stories during the six-week period. For this experiment, we took the 61,013 technical news and the associated user activity data from the dataset.

**MovieLens[37]:** This benchmark dataset for recommendations is frequently utilized. On a scale of 1 to 5, it contains roughly 1 million explicit ratings for movies from the MovieLens website. We translate ratings into implicit feedback, where each item is marked either with 1 or 0.

We use Microsoft Satori, a sizable commercial knowledge graph, to incorporate the additional knowledge data. We extract all triples in which the confidence of relations linked among entities is more than 0.8 by searching the neighbors of all entities in our news corpus in Microsoft Satori KG.

### 4.3 Baselines

To verify the effectiveness of our proposed method KGUPN，we use the following state-of-the-art methods as baselines:

•FM[38] is a benchmark decomposition model in which second-order feature interactions between inputs are considered.

•DKN[39] takes entity and word embeddings as channels and combines them in CNN for prediction.

• CKE[40] is a representative regularization-based method. It combines CF with diverse knowledge such as structural, textual, and visual knowledge in a recommendation framework.

• CFKG[41] transforms recommendation tasks into reasonable predictions of triplets, applying TransE to a unified graph including users, items and relationships.

• LibFM[42] is a feature-based decomposition model widely used in CTR scenarios. The inputs in this paper are the concatenated users, items and the corresponding average entity embeddings learned from TransR.

• RippleNet[43] combines regularization based and pathbased methods, which enrich representations by hopping to build relationship between items and user.

• GC-MC[44] is a model using GCN on graph-structured data, widely used in user-item bipartite graphs. This paper is applied to the user-item KG.

### 4.4 Experiment Setup

We choose all hyperparameters based on the results on the validation set. We split each dataset into a 6:2:2 train, evaluation, and test set. Each experiment was repeated 5 times and the average was taken as the final performance. By using the trained model, we select K items for users in the test set, which have the highest predicted click probability, with Recall@K and NDCG@K as evaluation metrics.

### 4.5 Performance Comparison

The performance comparison results are presented in Table II, and figures 3, 4, respectively. We have observations as following:

• KGUPN consistently yields the best performance on all the datasets. KGUPN improves over the strongest baselines as recall@20 by 4.6%, 4.69% in Mind and MovieLens, respectively. KGUPN effectively increases the recommendation accuracy by adding supplementary knowledge, user interaction information, and higher-order reasoning connectivity.

• FM and DKN achieve better performance than CFKG and CKE, indicating that the decomposition model can fully utilize item knowledge more than regularization-based methods. CFKG and CKE only use the embeddings of their aligned entities, while FM and DKN use the embeddings of connected entities to enrich the representation of items. In addition, CFKG and CKE keep high-order connections unchanged, while FM and DKN take their cross features as second-order connections between users and entities.

• The superior performance of LibFM compared to FM validates the importance of rich user representations, and it also points out the positive effects of correlation and neighbor modeling. However, libFM performs slightly better than GC-MC in Mind and performs worse in MovieLens. One possible reason is that the movie name is not very directional and short, which does not provide useful information.

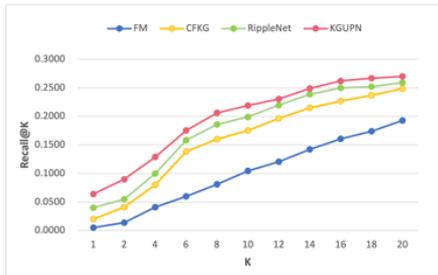

**Figure 3** Recall with top K on MIND datasets

Figure 3 and Figure 4 present the Recall and Hit Ratio with K on KGUPN and other baselines, FM, CFKG, RippleNet. We can observe the curve of KGUPN is consistently above the baselines as the K growing, which strongly proves the competitiveness of KGUPN.

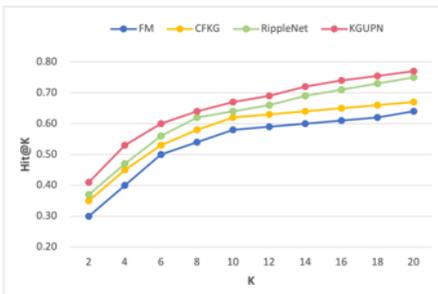

**Figure 4** Hit ratio with top K on MIND datasets

Also, we conducted experiments to quest the effect of hops on recommend performance, and the result is shown in Table III. It can be observed that a larger number of hops hardly improves performance but does incur heavier computational overhead on both datasets according to experiment results. Therefore, we set the hop number as k=3 for cost-effective.

**Table III** Effect of propagating hops in KGUPN

| Hop Num | Mind NDCG@10 | MovieLens NDCG@10 |
| --- | --- | --- |
| 1 | 0.1842 | 0.0732 |
| 2 | 0.2133 | 0.0817 |
| 3 | 0.2278 | 0.0943 |
| 4 | 0.2269 | 0.0945 |

**4.6 Ablation Study**

Next, we performed ablation experiments to verify the effectiveness of each layer in KGUPN. There are collaborative relations layer, propagation representation layer, and contextual information layer in KGUPN, we remove one of these layers each time and observe performance on dataset Mind and MovieLens. Results are shown in Table IV and the findings are as following:

**Table IV** Effect of layers in KGUPN

| Model | MIND | |
| --- | --- | --- |
| | Recall@20 | NDCG@10 |
| KGUPN | 0.2166 | 0.2278 |
| w/o Collaborative Relation | 0.2155 | 0.2262 |
| w/o Propagating | 0.2144 | 0.2242 |
| w/o Contexual Information | 0.2146 | 0.2249 |

• In MIND and MovieLens, the lack of collaborative relation layer leads to 0.51%, 1.31% reduction in Recall@20 separately, and have 0.68%, 1.63% reduction in NDCG@10.

• The remove of propagation representation layer resulting in 1.02%, 1.40% drop in Recall@20, and 1.56% ,2.43% drop in NDCG@10.

• The absence of contextual information layer resulting in 0.92%, 1.65% drop in Recall@20, and 1.26%,2.27% drop in NDCG@10.

We noticed that the absence of either one layer in KGUPN will cause a notable drop of performance, thus, all layers are necessary.

**4.7 Performance with respect to Epoch**

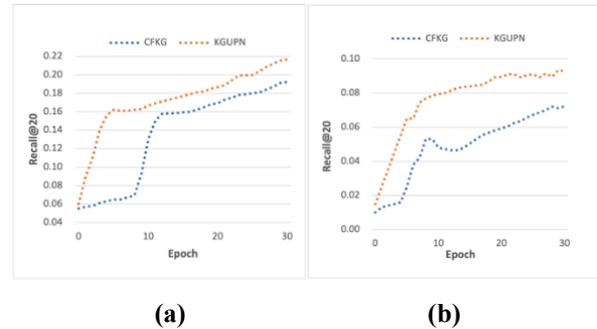

(a)  (b)

**Figure 5** Performance of each epoch of KGUPN and CGKG on Mind (a) and MovieLens (b)

Figure 5 is the performance of recall per epoch for CFKG and KGUPN. From the figure, we can find that KGUPN shows faster convergence than CFKG on MIND and MovieLens datasets, one of the reasons is because indirectly connected users and items are involved in optimizing mini-batch interaction pairs. Such convergence speed proves that KGUPN has better model capacity and is more effective in performing embedding propagation in the embedding space.

**5 Conclusions**

In this paper, we proposed KGUPN, an end-to-end framework that incorporates knowledge graph and user

*Proceedings of CCIS2022*awareness into scientific and technological news recommendation systems, solves the shortcomings of previous embedding-based and path-based knowledge graph-aware recommendation approaches. The propagation representation layer, the contextual information layer, and the collaborative relation layer are the three key layers that make up KGUPN. We carry out extensive experiments on two recommendation datasets. The results show that KGUPN significantly outperforms the other baselines.

## Acknowledgements

This work was supported by the National Natural Science Foundation of China (No.62192784, No.62172056).## References

[1] Feifei Kou, Junping Du, Congxian Yang, Yansong Shi, Wanqiu Cui, Meiyu Liang, and Yue Geng. Hashtag recommendation based on multi-features of microblogs. Journal of Computer Science and Technology, 2018, 33(4): 711-726.

[2] Ang Li, Junping Du, Feifei Kou, Zhe Xue, Xin Xu, Mingying Xu, Yang Jiang. Scientific and Technological Information Oriented Semantics-adversarial and Media-adversarial Cross-media Retrieval. arXiv preprint arXiv:2203.08615, 2022.

[3] Das, Abhinandan S., Mayur Datar, and Ashutosh Garg. "Google News Personalization." Proceedings of the 16th international conference on World Wide Web - WWW '07, 2007, 56(1):271–280.

[4] Wang, Chong, and David M. Blei. "Collaborative Topic Modeling for Recommending Scientific Articles." Proceedings of the 17th ACM SIGKDD international conference on Knowledge discovery and data mining - KDD '11, 2011, 488–56.

[5] Wenling Li, Yingmin Jia, and Junping Du. Distributed extended Kalman filter with nonlinear consensus estimate. Journal of the Franklin Institute, 2017, 354(17): 7983-7995.

[6] Yawen Li, Isabella Yunfei Zeng, Ziheng Niu, Jiahao Shi, Ziyang Wang and Zeli Guan, Predicting vehicle fuel consumption based on multi-view deep neural network, Neurocomputing, 502:140-147, 2022.

[7] Zeli Guan, Yawen Li, Zhe Xue, Yuxin Liu, Hongrui Gao, Yingxia Shao. Federated Graph Neural Network for Cross-graph Node Classification. In 2021 IEEE 7th International Conference on Cloud Computing and Intelligent Systems, 418-422, 2021.

[8] Joseph, Kevin, and Hui Jiang. "Content Based News Recommendation via Shortest Entity Distance over Knowledge Graphs." Companion Proceedings of The 2019 World Wide Web Conference, 2019, 61－72.

[9] Huang, Po-Sen, Xiaodong He, Jianfeng Gao, Li Deng, Alex Acero, and Larry Heck. "Learning Deep Structured Semantic Models for Web Search Using Clickthrough Data." Proceedings of the 22nd ACM International Conference on Information & Knowledge Management. 2013: 2333－2338.

[10] Wenling Li, Jian Sun, Yingmin Jia, Junping Du, and Xiaoyan Fu. Variance-constrained state estimation for nonlinear complex networks with uncertain coupling strength. Digital Signal Processing, 2017, 67: 107-115.

[11] Meguebli, Youssef, Mouna Kacimi, Bich-liên Doan, and Fabrice Popineau. "Stories around You - a Two-Stage Personalized News Recommendation." Proceedings of the International Conference on Knowledge Discovery and Information Retrieval, 2014. 125－134.

[12] Wenling Li, Yingmin Jia, Junping Du. Distributed consensus extended Kalman filter: a variance-constrained approach. IET Control Theory & Applications, 11(3): 382-389, 2017.

[13] Lian, Jianxun, Fuzheng Zhang, Xing Xie, and Guangzhong Sun. "Towards Better Representation Learning for Personalized News Recommendation: A Multi-Channel Deep Fusion Approach." Proceedings of the Twenty-Seventh International Joint Conference on Artificial Intelligence, 2018, 3805–3811

[14] Wenling Li, Yingmin Jia, and Junping Du. Tobit Kalman filter with time-correlated multiplicative measurement noise. IET Control Theory & Applications, 2016, 11(1): 122-128.

[15] Okura, Shumpei, Yukihiro Tagami, Shingo Ono, and Akira Tajima. "Embedding-Based News Recommendation for Millions of Users." Proceedings of the 23rd ACM SIGKDD International Conference on Knowledge Discovery and Data Mining, 2017. https://doi.org/10.1145/3097983.3098108.

[16] Hongwei Wang, Fuzheng Zhang, Xing Xie, and Minyi Guo. 2018. DKN: Deep Knowledge-Aware Network for News Recommendation. In Proceedings of the 2018 World Wide Web Conference on World Wide Web, WWW 2018, Lyon, France, April 23-27, 2018, Pierre-Antoine Champin, Fabien L. Gandon, Mounia Lalmas, and Panagiotis G.

[17] Qingping Li, Junping Du, Fuzhao Song, Chao Wang, Honggang Liu, Cheng Lu. Region-based multi-focus image fusion using the local spatial frequency. 2013 25th Chinese control and decision conference (CCDC), 2013: 3792-3796.

[18] Kim, Yoon. "Convolutional Neural Networks for Sentence Classification." Proceedings of the 2014 Conference on Empirical Methods in Natural Language Processing (EMNLP), 2014. https://doi.org/10.3115/v1/d14-1181.

[19] Deyuan Meng, Yingmin Jia, and Junping Du. Robust iterative learning protocols for finite-time consensus of multi-agent systems with interval uncertain topologies. International Journal of Systems Science, 2015, 46(5): 857-871.

[20] Hu, Binbin, Chuan Shi, Wayne Xin Zhao, et al. "Leveraging Meta-Path Based Context for Top- n Recommendation with a Neural Co-Attention Model." Proceedings of the 24th ACM SIGKDD International Conference on Knowledge Discovery & Data Mining, 2018. https://doi.org/10.1145/3219819.3219965.

[21] Mingxing Li, Yinmin Jia, and Junping Du. LPV control with decoupling performance of 4WS vehicles under velocity-varying motion. IEEE Transactions on Control Systems Technology 2014, 22(5): 1708-1724.

[22] Wang, Xiang, Dingxian Wang, Canran Xu et al. Explainable Reasoning over Knowledge Graphs for Recommendation. Proceedings of The AAAI Conference on Artificial Intelligence. 2019, 33(1): 5329-5336.

[23] Deyuan Meng, Yingmin Jia, Junping Du, and Fashan Yu, Tracking Algorithms for Multiagent Systems, In IEEE Transactions on Neural Networks and Learning Systems, 2013, 24(10): 1660-1676.Liang Xu, Junping Du, Qingping Li. Image fusion based on nonsubsampled contourlet transform and saliency-motivated pulse coupled neural networks. Mathematical Problems in Engineering, 2013.

[24] Lin, Yankai, Zhiyuan Liu, Maosong Sun. Learning Entity and Relation Embeddings for Knowledge Graph

Completion[C]// Proceedings of The AAAI Conference on Artificial Intelligence. 2019: 673-681.
[25] Deyuan Meng, Yingmin Jia, and Junping Du. Consensus seeking via iterative learning for multi-agent systems with switching topologies and communication time-delays. *International Journal of Robust and Nonlinear Control*, 2016, 26(17): 3772-3790.
[26] Peng Lin, Yingmin Jia, Junping Du, Fashan Yu. Average consensus for networks of continuous-time agents with delayed information and jointly-connected topologies. 2009 American Control Conference, 2009: 3884-3889.
[27] Wang, Hongwei, Miao Zhao, Xing Xie et al. "Knowledge Graph Convolutional Networks for Recommender Systems." The World Wide Web Conference on - WWW '19, 2019. https://doi.org/10.1145/3308558.3313417.
[28] Zeyu Liang, Junping Du, and Chaoyang Li. Abstractive social media text summarization using selective reinforced Seq2Seq attention model. Neurocomputing, 410 (2020): 432-440.
[29] Qing Ye, Chang-Yu Hsieh, Ziyi Yang, Yu Kang, Jiming Chen, Dongsheng Cao. A unified drug–target interaction prediction framework based on knowledge graph and recommendation system. Nat Commun 12, 6775 (2021). https://doi.org/10.1038/s41467-021-27137-3.
[30] Xinlei Wei, Junping Du, Meiyu Liang, and Lingfei Ye. Boosting deep attribute learning via support vector regression for fast moving crowd counting. Pattern Recognition Letters, 2019, 119: 12-23.
[31] Xiang Wang, Xiangnan He, Yixin Cao, et al. KGAT: Knowledge Graph Attention Network for Recommendation. 2019: 950–958.
[32] Jianghai Lv, Yawen Li, Junping Du, Lei Shi. E-Product Recommendation Algorithm Based on Knowledge Graph and Collaborative Filtering. Chinese Intelligent Systems Conference, 38-47, 2020.
[33] Yingxia Shao, Shiyue Huang, Yawen Li, Xupeng Miao, Bin Cui, Lei Chen. Memory-aware framework for fast and scalable second-order random walk over billion-edge natural graphs. The VLDB Journal, 30(5), 769-797, 2021.
[34] Jizhou Huang, Haifeng Wang, Yibo Sun, Miao Fan, Zhengjie Huang, Chunyuan Yuan, Yawen Li. HGAMN: Heterogeneous Graph Attention Matching Network for Multilingual POI Retrieval at Baidu Maps. Proceedings of the 27th ACM SIGKDD Conference on Knowledge Discovery & Data Mining. 3032-3040, 2021.
[35] Rendle, Steffen, Walid Krichene, Li Zhang, and John Anderson. "Neural Collaborative Filtering vs. Matrix Factorization Revisited." Fourteenth ACM Conference on Recommender Systems, 2020. https://doi.org/10.1145/3383313.3412488.
[36] Wu, Fangzhao, Ying Qiao, Jiun-Hung Chen, Chuhan Wu, Tao Qi, Jianxun Lian, Danyang Liu, et al. "Mind: A Large-Scale Dataset for News Recommendation." Proceedings of the 58th Annual Meeting of the Association for Computational Linguistics, 2020. https://doi.org/10.18653/v1/2020.acl-main.331.
[37] Harper, F. Maxwell, and Joseph A. Konstan. "The Movielens Datasets." ACM Transactions on Interactive Intelligent Systems 5, no. 4 (2016): 1–19. https://doi.org/10.1145/2827872.
[38] Steffen Rendle, Zeno Gantner, Christoph Freudenthaler, and Lars Schmidt-Thieme. Fast context-aware recommendations with factorization machines. Proceedings of the 34th international ACM SIGIR conference on Research and development in Information - SIGIR '11. 2011, 635–644
[39] Hongwei Wang, Fuzheng Zhang, Xing Xie, and Minyi Guo. 2018. DKN: Deep Knowledge-Aware Network for News Recommendation. In Proceedings of the World Wide Web Conference on World Wide Web. 2018, 1835–1844.
[40] Fuzheng Zhang, Nicholas Yuan, Defu Lian et al. Collaborative knowledge base embedding for recommender systems. In Proceedings of the 22nd ACM SIGKDD International Conference on Knowledge Discovery and Data Mining. 2016, 353–362.
[41] Qingyao Ai, Vahid Azizi, Xu Chen et al. Learning Heterogeneous Knowledge Base Embeddings for Explainable Recommendation. Algorithms. 2018:11 (9), 137.
[42] Steffen Rendle. Factorization Machines with LibFM. ACM Transactions on Intelligent Systems and Technology (TIST) 2012:3(3), 57.
[43] Hongwei Wang, Fuzheng Zhang, Jialin Wang et al. RippleNet: Propagating User Preferences on the Knowledge Graph for Recommender Systems. 2018, 417–426.
[44] Rianne van den Berg, Thomas N. Kipf, and Max Welling. Graph Convolutional Matrix Completion. In KDD. 2017.